\documentclass[twoside]{article}
\usepackage{qic,epsfig}

\renewcommand{\thefootnote}{\fnsymbol{footnote}}  %use symbolic footnote

\begin{document}
\setlength{\textheight}{8.0truein}    %FOR 2ND PAGE ONWARDS

\runninghead{
Quantum algorithm design using dynamic learning}
% UANTUM ALGORITHM DESIGN USING DYNAMIC LEARNING}
            {E.C. Behrman, J.E. Steck, P. Kumar, and K.A. Walsh}

\normalsize\textlineskip
\thispagestyle{empty}
\setcounter{page}{12}

%\copyrightheading{Vol.}{No.}{Year}{Page Nos.}
\copyrightheading{8}{1\&2}{2008}{0012--0029}

\vspace*{0.88truein}

\alphfootnote

\fpage{12}

\centerline{\bf QUANTUM ALGORITHM DESIGN USING DYNAMIC LEARNING}
\vspace*{0.37truein}
\centerline{\footnotesize E.C. BEHRMAN}
\vspace*{0.015truein}
\centerline{\footnotesize\it Department of Physics, Wichita State University}
\baselineskip=10pt
\centerline{\footnotesize\it Wichita, KS 67260-0032,USA}
\vspace*{10pt}
\centerline{\footnotesize J.E. STECK}
\vspace*{0.015truein}
\centerline{\footnotesize\it Department of Aerospace Engineering, Wichita State University}
\baselineskip=10pt
\centerline{\footnotesize\it Wichita, KS 67260-0044, USA}
\vspace*{10pt}
\centerline{\footnotesize P. KUMAR}
\vspace*{0.015truein}
\centerline{\footnotesize\it Department of Electrical and Computer Engineering, Wichita State University}
\baselineskip=10pt
\centerline{\footnotesize\it Wichita, KS 67260-0044, USA}
\vspace*{10pt}
\centerline{\footnotesize K.A. WALSH}
\vspace*{0.015truein}
\centerline{\footnotesize\it Department of Physics, Wichita State University}
\baselineskip=10pt
\centerline{\footnotesize\it Wichita, KS 67260-0032, USA}
\vspace*{0.225truein}
\publisher{15 October 2006}{7 August 2007}%\vspace*{10pt}
\vspace*{0.21truein}

\abstracts{We present a dynamic learning paradigm for ``programming'' a general quantum computer.  A 
learning algorithm is used to find the control parameters for a coupled qubit system, such 
that the system at an initial time evolves to a state in which a given measurement corresponds 
to the desired operation. This can be thought of as a quantum neural network.   We first apply 
the method to a system of two coupled superconducting quantum interference devices (SQUIDs), and 
demonstrate learning of both the classical gates XOR and XNOR. Training of the phase produces a 
gate congruent to the CNOT modulo a phase shift.  
Striking out for somewhat more interesting territory, we attempt learning 
of an entanglement witness for a two qubit system.  Simulation shows a reasonably successful 
mapping of the entanglement at the initial time onto the correlation function at the final time 
for both pure and mixed states.  For pure states this mapping requires knowledge of the phase 
relation between the two parts; however, given that knowledge, this method can be used to measure 
the entanglement of an otherwise unknown state.  The method is easily extended to multiple qubits 
or to quNits.
}{}{}

\vspace*{10pt}
\vspace*{3pt}
\communicate{S Braunstein~\&~B Terhal}

\noindent
\vspace*{1pt}\textlineskip      %) USE THIS MEASUREMENT WHEN THERE IS
\section{Introduction}          %) A SECTION HEADING
\vspace*{-0.5pt}
\noindent
 Recently there has been growing interest in quantum computing \cite{nielsen, berman}. The possibilities seem vast. 
Beyond the improvements in size and speed, is the ability, at least in principle, to do classically impossible 
calculations. Two aspects of quantum computing make this possible: quantum parallelism and entanglement. While a 
computational setup can be constructed \cite{behrman} which makes use of quantum parallelism (superposition) only, 
use and manipulation of entanglement as well realizes the full power of quantum computing and 
communication \cite{bennett, div, linden}. 
  
The major bottleneck to the use of quantum computers, once they are designed and built, is the paucity of algorithms 
that can make use of their power. At present, there are only a few major algorithms: Shor's factorization \cite{shor},  
Grover's data base search \cite{grover}, the Jones polynomial approximation \cite{aharonov}. It is not yet at all clear 
that a way will be or can be found to generate algorithms efficiently to solve general problems on quantum computers, 
as pointed out by Nielsen\cite{nielsen2}, though some recent work\cite{nielsen2, khaneja} using a geometric approach 
may prove fruitful.
  
In previous work \cite{behrman}, we have proposed the use of quantum adaptive computers to answer this need. An adaptive 
computer, since it can be trained, adapts to learn and in a sense constructs its own algorithm for the problem from the 
training set supplied. A quantum neural computer shows promise for constructing algorithms to solve problems that are 
inherently quantum mechanical. Here, we develop a dynamic learning algorithm for training a quantum computer and 
demonstrate successful learning of some simple benchmark applications.  In addition, we show that this method can be 
used for learning of an entanglement witness\cite{filip} for an input state.  We show that our witness approximately 
reproduces the entanglement of formation for large classes of states.  Generalization to systems of more than two 
qubits \cite{greenberger}, or to multiple level systems \cite{kasz}, is straightforward.

\vspace*{1pt}\textlineskip      %) USE THIS MEASUREMENT WHEN THERE IS
\section{ Coupled Two-qubit System: QNN}                %) A SECTION HEADING
\vspace*{-0.5pt}
\noindent
Two interacting qubits, labeled A and B, can be used to build a quantum gate where each qubit interacts with a 
coupling (connectivity) that can be externally adjusted. This is a dynamical system that can be prepared in an 
initial (input) state, which then evolves in time to the point where it can be measured at some final time to 
yield an output. Adjustable physical parameters of the qubits allow ``programming'' to ``compute'' a specified output 
in response to a given input. 
  
Consider a two-qubit quantum system that evolves in time according to the Hamiltonian:

\noindent
\begin{equation}
H = K_{A} \sigma_{xA} + K_{B} \sigma_{xB} + \varepsilon_{A} \sigma_{zA} + \varepsilon_{B} \sigma_{zB} + \zeta \sigma_{zA} \sigma_{zB} 
\label{eqone}
\end{equation}
where $\{ \sigma \}$ are the Pauli operators corresponding to each of the two qubits, 
A and B, $K_{A}$ and $K_{B}$ are the tunneling amplitudes, $\varepsilon_{A}$ and, 
$\varepsilon_{B}$  are the biases, and $\zeta$ the qubit-qubit coupling. This Hamiltonian 
can also be written \cite{yamamoto} as

\noindent
\begin{eqnarray}
H  = & \sum_{n_{1},n_{2} = 0,1} E_{n_{1},n_{2}} |n_{1},n_{2} \rangle \langle n_{1},n_{2} | & - \frac{E_{J1}}{2} \sum_{n_{2} = 0,1} (|0 \rangle \langle 1 | + |1 \rangle \langle 0|) \otimes |n_{2} \rangle \langle n_{2} | \nonumber \\ 
&   &   - \frac{E_{J2}}{2} \sum_{n_{1} = 0,1} |n_{1} \rangle \langle n_{1} |\otimes (|0 \rangle \langle 1 | + |1 \rangle \langle 0|)
%\label{eqtwo}
\end{eqnarray}
using the two-qubit charge basis $ |00\rangle, |10\rangle, |01\rangle, |11\rangle$. This could represent a number of 
different possible physical systems, {\it e.g.}, trapped ions \cite{gulde} or nuclear magnetic 
resonance \cite{vandersypen}.  Here we take as our physical model an electrostatically coupled two-SQUID 
system \cite{yamamoto, han}.  With appropriately timed pulses of the bias amplitudes $\varepsilon_{A}$ 
and $\varepsilon_{B}$ , entangled states can be prepared, and logic gates such as the CNOT or Toffoli 
reproduced \cite{yamamoto, gagnebin}.   While we consider here a SQUID system with the particular physical 
identifications, above, of the parameters in the Hamiltonian, the method is by no means limited to that physical 
implementation of a two-quibit system, or, indeed, to a two-qubit system.  It is easy to see how to generalize to 
any Hamiltonian containing appropriately adjustable parameters.

The density matrix of the system as a function of time obeys the Schrödinger equation 
$\frac{d \rho}{dt} = \frac{1}{i \hbar}[H, \rho] $ , whose formal solution is $\rho(t) = \exp (iLt) \rho (t)$.  
This is of similar mathematical form to the equation for information propagation in a neural network, as follows.  
For a traditional artificial neural network, the calculated activation $\phi_{i}$ of the $i^{th}$ neuron is performed 
on the signals $\{ \phi_{i} \}$ from the other neurons in the network, and is given by 
$\phi_{i} = \sum_{j}w_{ij} f_{j}(\phi_{j})$, where $w_{ij}$ is the weight of the connection from the output of neuron 
$j$ to neuron $i$,  and $f_{i}$ is a bounded differentiable real valued neuron activation function for neuron $j$.    
Individual neurons are connected together into a network to process information from a set of input neurons to a set 
of output neurons.  The network is an operator $F_{W}$ on the input vector $\phi_{input}$. $F_{W}$ depends on the 
neuron connectivity weight matrix $W$ and propagates the information forward to calculate an output vector 
$\phi_{output}$; that is, $\phi_{output}= F_{W} \phi_{input }$.  The time evolution equation for the density matrix 
maps the initial state (input) to the final state (output) in much the same way.  The parameters playing the role of 
the adjustable weights in the neural network are the set $\{K_{A}, K_{B}, \varepsilon_{A}, \varepsilon_{B}, \zeta \}$, 
all of which can be adjusted experimentally as functions of time for the SQUID system under consideration \cite{yamamoto, 
han}.   By adjusting the parameters using a neural network type learning algorithm we can train the system to evolve in 
time to a final state at the final time $t_{f}$ for which the desired measure has been mapped to the function we wish 
the net to compute: logic gates, or, since the time evolution is quantum mechanical, a quantum function like the 
entanglement. Indeed, if we think of the time evolution operator in terms of the Feynman path integral picture 
\cite{feynman}, the parallel becomes even more compelling: Instantaneous values taken by the quantum system at 
intermediate times, which are integrated over, play the role of ``virtual neurons'' \cite{behrman}.

In any case the real time evolution of the two-qubit system can be treated as a neural network, because its evolution 
is a nonlinear function of the various adjustable parameters (weights) of the Hamiltonian.  For as long as coherence 
can be maintained experimentally, it is a quantum neural network (QNN).  Thus, if we can find values for the parameters 
such that the set of $\{inputs, outputs \}$ matches a measure of entanglement, we can use this setup as an experimental  
means of measuring the entanglement of any prepared state of the system, whether that state is analytically known or 
unknown.  In this paper we find those parameters by training a simulation of the two-qubit system to a set of four 
input-output pairs.  We then test the simulated net on a large number of additional states.  The net is said to have 
generalized if the results on the testing set are correct and consistent.  We show that this is so.
\setcounter{footnote}{0}
\renewcommand{\thefootnote}{\alph{footnote}}

\vspace*{1pt}\textlineskip      %) USE THIS MEASUREMENT WHEN THERE IS
\section{ Dynamic Learning for the Quantum System}              %) A SECTION HEADING
\vspace*{-0.5pt}
\noindent
The goal of learning as applied to this quantum system is to control the system via the external parameters 
(tunneling, field and coupling values) to force it to calculate target outputs in response to given inputs. 
This is essentially a neural network supervised learning paradigm extended to the quantum system. The method, 
derived below, follows the methodology of Yann LeCun's Lagrangian formulation derivation of backpropagation \cite{lecun} 
and Paul Werbos's description of backpropagation through time \cite{werbos}, and follows some of our earlier 
work \cite{steck,behrman2,skinner} on learning in non-linear optical materials and in training of quantum Hopfield 
networks. Our method uses the density matrix representation for generality.
  
We derive a learning rule for the quantum system based on dynamic backpropagation for time dependent recurrent 
neural networks. Given an input (initial density matrix), $\rho (0)$, and a target output, $d$, a training pair 
from a training set, we want to develop a weight update rule based on gradient descent to adjust the system parameters 
(tunneling, field and coupling), {\it i.e.}, train the system ``weights'', to reduce the squared error between the 
target, $d$, and the output, $Output$. While training the weights, the system's density matrix, $\rho(t)$, 
is constrained to satisfy the Schrödinger equation for all time in the interval $(0,t_{f})$.
  
  We define a Lagrangian, $L$, to be minimized as
\noindent
\begin{equation}
L = \frac{1}{2} [d - \langle O(t_{f}) \rangle ]^{2} + \int_{0}^{t_{f}} \lambda^{+}(t)(\frac{\partial \rho}{\partial t} + \frac{i}{\hbar}[H,\rho])\gamma(t) dt
\label{eqthree}
\end{equation}
where the Lagrange multiplier vectors are $\lambda^{+}(t)$ and $\gamma (t)$  (row and column, respectively),  
and $O$ is an output measure (or some function of a measure), which can be specified for the particular problem 
under consideration and is defined as:
\noindent
\begin{equation}
Output = \langle O(t_{f}) \rangle = tr[\rho(t_{f})O] = \sum_{i} p_{i} |\psi_{i}(t_{f})\rangle \langle \psi_{i}(t_{f})|O = 
\sum_{i} p_{i} \langle \psi_{i}(t_{f})|O|\psi_{i}(t_{f})\rangle 
\label{eqfour}
\end{equation}
where $tr$ stands for the trace of the matrix. We take the first variation of $L$ with respect to $\rho$, 
set it equal to zero, then integrate by parts to give the following equation which can be used to calculate 
the vector elements of the Lagrange multipliers that will be used in the learning rule:
\noindent
\begin{equation}
\gamma_{i} \frac{\partial \gamma_{j}}{\partial t} + \frac{\partial \lambda_{i}}{\partial t} \gamma_{j} - \frac{i}{\hbar} \sum_{k} \lambda_{k} H_{ki} \gamma_{j} + \frac{i}{\hbar} \sum_{k} \lambda_{i} H_{jk} \gamma_{k} = 0
\label{eqfive}
\end{equation}
with the boundary conditions at the final time $t_{f}$  given by
\noindent
\begin{equation}
-[d-\langle O(t_{f}) \rangle ] O_{ji} + \lambda_{i}(t_{f})\gamma_{j}(t_{f}) = 0
\label{eqsix}
\end{equation}

The gradient descent learning rule is given by 
\noindent
\begin{equation}
w_{new} = w_{old} - \eta \frac{\partial L}{\partial w}
\label{eqseven}
\end{equation}
for each weight parameter $w$, where $\eta$ is the learning rate and
\noindent
\begin{equation}
\frac{\partial L}{\partial w} = \frac{i}{\hbar} \int_{0}^{t_{f}} \lambda^{+}(t) [\frac{\partial H}{\partial w},\rho] \gamma(t) dt = \frac{i}{\hbar} \int_{0}^{t_{f}}\sum_{ijk} (\lambda_{i}(t) \frac{\partial H_{ik}}{\partial w} \rho_{kj} \gamma_{j} - \lambda_{i}(t) \rho_{ik} \frac{\partial H_{kj}}{\partial w} \gamma_{j} ) dt
\label{eqeight}
\end{equation}

Note that because of the Hermiticity of the Hamiltonian, $H$, and the density matrix 
$\rho$, $\lambda_{i} \gamma_{j} = \lambda_{j} \gamma_{i} $ and the derivative of the 
Lagrangian, $L$, with respect to the weight, $w$, as given by (8), will be a real number.  
This simplifies the calculation somewhat.
  
  The time evolution of the quantum system is calculated by integrating the Schrödinger equation in MATLAB 
Simulink \cite{matlab}. The ODE4 fixed step size solver was used with a step size of 0.05 ns. Discretization 
error for the numerical integration was checked by redoing the calculations with a timestep of a tenth the size; 
results were not affected.  Since the error needs to be back propagated through time, the integration has to be 
carried out from $t_{f}$ to 0.  To implement this in MATLAB Simulink, a change of variable is made by 
letting $t' = t_{f}-t$.  Instead of using Simulink, the Schrödinger wave equation can also be integrated with any 
standard numerical method such as a FORTRAN or C-code program, which we have also done, to validate the MATLAB simulation.

\vspace*{1pt}\textlineskip      %) USE THIS MEASUREMENT WHEN THERE IS
\section{ Benchmark Training Results for Two-qubit Gates} %) A SECTION HEADING
\vspace*{-0.5pt}
\noindent
  We now train a two-SQUID quantum system in simulation, to produce two-input one-output classical logic gates, 
specifically the XOR and XNOR. The system is initialized to (prepared in) the input states shown in Table 1 and 
allowed to evolve for 300 ns. We choose as the output measure $O$ the state of the second qubit, B, at the final time 
({\it i.e.}, $O = \sigma_{zB}(t_{f}).$). We call B the ``target'' qubit (while A is the ``control'' qubit.)  
The measure is applied to find the state of the system at this final time $t_{f}$ and compared to the target 
output for that particular input. The parameters $\lambda$ and $\gamma$ are initialized with this error according 
to (6), and this is dynamically back propagated through time according to (5) and the weights updated according 
to (7). Since the control qubit A does not change its state and we are only concerned with measuring the output 
state of the target qubit B at the final time, we do not need to train the weights corresponding to the parameters 
for the control SQUID A, {\it i.e.}, $K_{A}$ and $\varepsilon_{A}$ need not be trained. Thus, we only train the 
weights corresponding to the parameter values $K_{B}$, $\varepsilon_{B}$ and $\zeta_{AB}$. The value 
for $\varepsilon_{A}$ can be chosen to be any arbitrarily high value, say 1.0 GHz; this maintains the control 
qubit A in its original state. The initial values of the parameters (weights) before training are shown in 
Tables 2 and 3 for the XOR and XNOR gates, respectively, along with the final trained parameters (weights) 
for each logic gate. The field $\varepsilon_{B}$ is allowed to vary with time. The simulation models this by 
allowing the field to vary every 100 ns, as a series of step functions: $\varepsilon_{B}(1)$ and $\varepsilon_{B}(2)$. 
The trained responses are shown in Table 1 along with the RMS error for each logic gate and the number of 
epochs of training used for each.  The trained parameters agree with our previous analytic results\cite{gagnebin}. 
Note that the parameter values can be rescaled overall.

\vspace*{4pt}   %only when needed
\begin{table}[hb]
\tcaption{ Training data for  two input one output  logic gates.}
\centerline{\footnotesize\smalllineskip
\begin{tabular}{l c c c c }\\
Input &  \multicolumn{2}{c}{XOR}   &  \multicolumn{2}{c}{XNOR}  \\ \cline{2-5}
$|\psi\rangle$                     &    Target & Output & Target & Output \\
\hline \\
$|00\rangle$  &   -1           &   -0.9919 & +1     & 0.9902 \\
$|01\rangle$  &   +1           &   0.9920 & -1     & -0.9903 \\
$|10\rangle$  &   +1           &   0.9902 & -1     & -0.9919 \\
$|11\rangle$  &   -1           &   -0.9903 & +1     & 0.9920 \\
\hline  \\
RMS   &  \multicolumn{2}{c}{0.00446}   &   \multicolumn{2}{c}{0.00447}   \\
Epochs   &   \multicolumn{4}{c}{300}   \\
\hline  \hline \\
\end{tabular}}
\end{table}

\vspace*{4pt}   %only when needed
\begin{table}[hb]
\tcaption{ Initial and trained parameters (weights) for XOR gate, in MHz.}
\centerline{\footnotesize\smalllineskip
\begin{tabular}{c c c }\\
\hline
\hline
Parameter (MHz) & Initial & XOR-trained \\
\hline
$K_{A}$  & 2.1333 & 2.1333 \\
$K_{B}$  & 2.1333 &  1.2684 \\
$\zeta_{AB}$  & 0.1 & -0.97981 \\
$\varepsilon_{A}(1) = \varepsilon_{A}(2)$  & 1,000 &  1,000 \\
$\varepsilon_{B}(2)$  & 0.1 &  1.0518 \\
$\varepsilon_{B}(3)$  & 0.1 &  1.0534 \\
\hline \hline \\
\end{tabular}}
\end{table}

\vspace*{4pt}   %only when needed
\begin{table}[hb]
\tcaption{Initial and trained parameters (weights) for XNOR gate, in MHz.}
\centerline{\footnotesize\smalllineskip
\begin{tabular}{c c c }\\
\hline
\hline
Parameter (MHz) & Initial & XNOR-trained \\
\hline
$K_{A}$  & 2.1333 & 2.1333 \\
$K_{B}$  & 2.1333 &  1.2682 \\
$\zeta_{AB}$  & 0.1 & 0.97973 \\
$\varepsilon_{A}(1) = \varepsilon_{A}(2)$  & 1,000 &  1,000 \\
$\varepsilon_{B}(2)$  & 0.1 &  1.0508 \\
$\varepsilon_{B}(3)$  & 0.1 &  1.0524 \\
\hline \hline \\
\end{tabular}}
\end{table}

\vspace*{1pt}\textlineskip      %) USE THIS MEASUREMENT WHEN THERE IS
\section{ Quantum Control} %) A SECTION HEADING
\vspace*{-0.5pt}
\noindent
  It should be noted that our chosen measurement, $O = \sigma_{zB}(t_{f})$, does not check the phase of the
 final state.  That is, Tables 1-3 show only the acquisition of the {\em classical} gates XOR and XNOR.  In order 
to show that we have a {\em quantum} gate we need to train the actual output state, directly.  That is, 
given an input state, we wish to adjust the parameters of the system such that it evolves to another given state.  
This is also known as ``quantum control'' \cite{shapiro}.  Of course, we cannot use the density matrix approach if 
we wish to do this.  A similar training procedure using kets is as follows. The output of the quantum system is taken 
as the overlap of the state of the system at the final time, $t_{f}$, with the desired state:
\noindent
\begin{equation}
Out = \langle \psi_{desired} | \psi(t_{f}) \rangle
\,. \label{eqnine}
\end{equation}

We define a Lagrangian, $L$, to be minimized as
\noindent
\begin{equation}
L = \frac{1}{2} |1-\langle \psi_{desired} | \psi(t_{f}) \rangle |^{2} + \int_{0}^{t_{f}} \lambda^{*} [\frac{d|\psi\rangle}{dt}-\frac{1}{i \hbar} H |\psi \rangle ] dt + \int_{0}^{t_{f}}[\frac{d|\psi\rangle}{dt}-\frac{1}{i \hbar} H |\psi \rangle ]^{*} \lambda dt
\,, \label{eqten}
\end{equation}
where * indicates the complex conjugate.  To minimize $L$, we set the first variation of  $L$ with 
respect to $ | \psi \rangle$ equal to zero. After integration by parts, this gives a differential equation 
for $\lambda$ as
\noindent
\begin{equation}
\frac{d \lambda}{dt} = \frac{1}{i \hbar} H^{*} \lambda
\,, \label{eqeleven}
\end{equation}
with a condition at the final time as
\noindent
\begin{equation}
\lambda^{*}(t_{f}) = {\rm Re}[(1-\langle \psi(t_{f})|\psi_{des}\rangle )\langle \psi_{des} |]
\,. \label{eqtwelve}
\end{equation}
  
Again, we can solve this equation backward in time, and take the variation of $L$ with respect to a weight parameter, 
$w$, where $w$ represents any one of the parameters from the Hamiltonian, $H$, such as $K_{A}$, $K_{B}$, etc., to 
give the parameter (weight) update learning rule (7), where now
\noindent
\begin{equation}
\frac{\partial L}{\partial w} = \int_{0}^{t_{f}} 2 {\rm Re}[\frac{1}{i \hbar} \lambda^{*} \frac{\partial H}{\partial w} | \psi \rangle ] dt
\,, \label{eqthirteen}
\end{equation}
and $\eta$ is again the learning rate.  Training using this method is less stable than with the density matrix; 
in particular, if we try to pin the control  qubit A as before, with a large value of $\varepsilon_{A}$, there are 
rapid oscillations in the phase which make computation difficult.  Fortunately one can get the same pinning result 
by setting $K_{A} = 0$.  (Use of the density matrix method gets rid of this phase oscillation problem but of course 
does not allow the user to determine the phase of the output.) 

Results are shown in Table 4.  This is not the CNOT, since one of the qubits has an extra phase\cite{divincenzo}
 of $\pi$; however, this gate, followed by the ``controlled Z'' gives the CNOT. 
(The solution for single-qubit phase shifts gates is 
obvious: set $\zeta_{AB} = 0$ , and let $\varepsilon_{B}$ be nonzero for the amount of time necessary.)  
   
It should be noted that this, like the application in the previous section, is a ``blind'' application of our method.  
Recent work \cite{khaneja} especially by Khaneja, {\it et al.}, using a geometric approach, shows that careful analysis 
can produce a much more optimal (efficient) realization of a quantum operator.  Still, our method allows relatively 
quick learning, and can be readily applied even when the problem -­or operator --is not well understood, as with the 
entanglement witness in the next section.

\vspace*{4pt}   %only when needed
\begin{table}[hb]
\tcaption{Training of the CNOT.}
\centerline{\footnotesize\smalllineskip
\begin{tabular}{l c c c c} \hline \hline
Input                    &   \multicolumn{4}{c }{Output amplitudes} \\ \cline{2-5}
$|\psi \rangle$      &    $|00\rangle$    &   $|01\rangle$     &     $|10\rangle$    &   $|11\rangle$   \\
\hline \\
$|00\rangle$    &   $0.9998 + 0.009i$   &  $0.0006 + 0.0188i$   &   $0.0000$          &    $0.0000$   \\
$|01\rangle$    &   $0.0006 + 0.0188i$  &  $0.9977 -0.0653i$   &   $0.0000$          &    $0.0000$   \\
$|10\rangle$    &   $0.0000$            &   $0.0000$            & $0.0323 + 0.0i$     &   $-0.9993 + 0.0202i$ \\
$|11\rangle$    &   $0.0000$            &    $0.0000$           & $0.9993 + 0.0203i$  &  $0.0292 -0.0085i$ \\
\hline
\multicolumn{5}{c}{$K_{A} = 0.0 = \varepsilon_{A}(1) =\varepsilon_{A}(2)$; $\zeta_{AB} = 0.0096$} \\
\multicolumn{5}{c}{$K_{B} = 0.0054$; $\varepsilon_{B}=(-0.0606, -0.0133, -0.0055)$} \\
\hline \hline 
\end{tabular}}
\end{table}

\vspace*{1pt}\textlineskip      %) USE THIS MEASUREMENT WHEN THERE IS
\section{ Entanglement Witness} %) A SECTION HEADING
\vspace*{-0.5pt}
\noindent

  Of course it is known how to produce simple universal gates with any of a number of physical implementations, 
and as long as we know how to decompose our desired computation into those simpler blocks we do not need to do it 
``all at once.''  Showing that our quantum neural network can be trained to do progressively more complicated gates 
does have some advantages: Online training automatically adjusts for small effects not taken into account in whatever 
model was used to design the algorithm, for example.  
  
  However of much more interest is the possibility that a neural or AI approach can help us calculate things we 
do not have an algorithm for, and/or which we do not know how to decompose into  simple gates.  Entanglement is a 
good example.  It is widely thought that the power of quantum computing and communications relies heavily on the 
use and manipulation of entanglement \cite{div,linden}.  But the quantification of entanglement is still not fully 
understood even for pure states, and for mixed states the situation is cloudier.  For many practical applications we 
will also need an extension to systems of more than two qubits \cite{greenberger} or to entangled pairs of N-state 
systems, ``quNits'' \cite{kasz}.  
  
  Two prominent universal measures do exist: those of Bennett {\it et al.}\cite{bennett2} and Wootters\cite{wootters}, 
and of Vedral {\it et al.}\cite{vedral}.  The first is the entanglement of formation, which for pure states is equal 
to the von Neumann entropy of the reduced single-qubit states of the two-qubit system.  The second constructs a space 
of density matrices, containing a subspace of unentangled states, and defines as the entanglement of a state the 
minimum distance of its density matrix to that subspace.  The two measures do not give the same answer for a number 
of states, but each is internally consistent.  We now use our training method to map an entanglement witness to a 
single (local) measure at the final time\cite{dynloc}.  Unlike the method of Vedral, it does not require a minimization 
procedure, which can become cumbersome especially if the number of qubits is large; unlike both, it can in principle 
easily be used experimentally to measure the entanglement of an unknown state.  We show that our method approximately 
reproduces the entanglement of formation for large classes of states.  
  
  We return to a two-qubit system, for which the Hamiltonian is given by (1).  We take as our output the square of 
the two-qubit correlation function at the final time, that is, 
$ \langle O(t_{f}) \rangle = \langle \sigma_{zA} (t_{f}) \sigma_{zB}(t_{f}) \rangle^{2} = [ tr (\rho (t_{f}) \sigma_{zA} \sigma_{zB})]^{2}$. 
(We use here the square of the correlation function so that the range of the output will be [0,1] for convenience; 
the only modification necessary in Eq. (8) is the multiplication by 
$2\langle \sigma_{zA}(t_{f}) \sigma_{zB}(t_{f}) \rangle$.)   In Table 5 we present a representative sample of the 
kinds of states whose entanglement we wish to calculate, along with the entanglement of each as calculated by the 
methods of references\cite{bennett2,wootters,vedral}.  The Bell and EPR states are maximally entangled, and product 
and (completely) mixed states are minimally entangled.  We chose a training set of four: one completely entangled 
state, one unentangled state, one classically correlated but unentangled state, and one partially entangled state.  
We used a time of evolution of 1000 $\hbar$, enough for the system to go through one complete oscillation. Results 
are presented in Table 6.  The RMS error for the set, after training, is essentially zero.  Table 7 lists the 
parameter values the system trained to in order to achieve the entanglement witness.  We then tested the witness 
on a testing set which included a maximally entangled state of a type not seen before by the net, the EPR state; 
a pure unentangled state; a correlated unentangled state, a different partially entangled state, and a mixed state.  
Results for the testing set are shown in Table 8.  The error for the testing set is also essentially zero.  
A large number of permutations of possible states have been tried with exactly similar results. The quantum neural 
net has learned to compute an approximate general measure of entanglement.   
  
\vspace*{4pt}   %only when needed
\begin{table}[htbp]
\tcaption{ Some possible states of the two-qubit system.  The relative amplitudes (for the ket states) are given 
without normalization for clarity.  The first two are maximally entangled. The second two are product states (
flat = $(|0 \rangle + |1 \rangle)_{A}(|0 \rangle + |1 \rangle)_{B}$ and 
$C = | 0 \rangle_{A}( | 0 \rangle + \gamma | 1 \rangle)_{B}$) and thus have zero entanglement; and $P$ is 
partially entangled. $M$ is a mixed state and cannot be expressed as a ket; its density matrix is given instead. 
The classical correlation is computed as $\langle \sigma_{zA}(0) \sigma_{zB}(0) \rangle$.  $C$ and $M$ are 
classically correlated but not entangled.  The last three columns show the entanglement as calculated by the 
methods of Bennett\cite{bennett2}  and Vedral \cite{vedral}, using the von Neumann metric and the Bures metric.  
Both distance measures have been normalized to unity.}
\centerline{\footnotesize\smalllineskip
\begin{tabular}{|l| c c c c |c| c c c |}\hline \\
State  &\multicolumn{4}{|l|}{Relative amplitudes of} & Classical &\multicolumn{3}{|l|}{Entanglement}\\ \cline{2-5} \cline{7-9}
{} & $|00\rangle$ & $|01\rangle$ &$|10\rangle$ &$|11\rangle$ & Correlation & Bennett & von Neumann & Bures \\
\hline \\
Bell  & 1 & 0 & 0 & $e^{i \theta}$ & 1 & 1 & 1 & 1 \\
EPR  & 0 & 1 & $e^{i \theta}$  & 0 & -1 & 1 & 1 & 1 \\
flat  & 1 & 1 & 1  & 1 & 0 & 0 & 0 & 0 \\
$C$ & 1 & $\gamma$ & 0  & 0 & $\frac{1-|\gamma|^{2}}{1+|\gamma|^{2}}$ & 0 & 0 & 0 \\
$P$  & 0 & 1 & 1  & 1 & $-1/3$ & 0.55 & 0.32 & 0.46 \\ \hline \cline{1-5}
$M$ & \multicolumn{4}{|c|}{$(|00 \rangle \langle 00| + |11 \rangle \langle 11|)/2$} & 1 & 0 & 0 & 0 \\ \hline 
\end{tabular}}
\end{table}

\vspace*{4pt}   %only when needed
\begin{table}[htbp]
\tcaption{Training data for QNN entanglement witness.}
\centerline{\footnotesize\smalllineskip
\begin{tabular}{l l l l}\\
\hline
Input state & Initial & Desired &Trained \\
\hline
Bell, $\delta = 0$ & 1.0 & 1.0 & 0.99997 \\
Flat                        & 0.0 & 0.0  &  $2.01 \times 10^{-6}$  \\
$C$, $\gamma = 0.5$ & 0.36 & 0.0  &  $2.61 \times 10^{-5}$  \\
$P$                             & 0.11  & 0.44317  & 0.44317  \\
\hline
RMS    &   { }  & $1.08 \times 10^{-5}$ & { }  \\
Epochs  &  { }   &  2000  &  {} \\
\hline \\
\end{tabular}}
\end{table}

\vspace*{4pt}   %only when needed
\begin{table}[htbp]
\tcaption{Initital and trained parameters for entanglement, in MHz.}
\centerline{\footnotesize\smalllineskip
\begin{tabular}{l c c }\\
\hline
Parameter (MHz)   &   Initial   &   Trained  \\
\hline
$K_{A}(1) $  &  2.5  &   2.3576 \\
$K_{A}(2) $  &  2.5   &  2.3576 \\
$K_{A}(3) $  &  2.5   &  2.3577  \\
$K_{A}(4) $  &  2.5   &  2.3461  \\
$K_{B}(1) $  &  2.5   &  2.3576 \\
$K_{B}(2) $  &  2.5   &  2.3576 \\
$K_{B}(3) $  &  2.5   &   2.3576 \\
$K_{B}(4) $  &  2.5   &   2.3546 \\
$\zeta(1)$        &  0.1    &  0.045026  \\
$\zeta(2)$        &  0.1    &  0.10117  \\
$\zeta(3)$        &  0.1    &  0.10771  \\
$\zeta(4)$        &  0.1    &  0.044221  \\
$\varepsilon_{A}(1)$   &   0.1  & 0.10913  \\
$\varepsilon_{A}(2)$   &   0.1  & 0.03768  \\
$\varepsilon_{A}(3)$   &   0.1  & 0.08671  \\
$\varepsilon_{A}(4)$   &   0.1  & 0.071464  \\
$\varepsilon_{B}(1)$   &   0.1  & 0.10913  \\
$\varepsilon_{B}(2)$   &   0.1  & 0.063774  \\
$\varepsilon_{B}(3)$   &   0.1  & 0.038802  \\
$\varepsilon_{B}(4)$   &   0.1  & 0.072387  \\
\hline \hline \\
\end{tabular}}
\end{table}

\vspace*{4pt}   %only when needed
\begin{table}[htbp]
\tcaption{ Representative testing results for the quantum neural network.  $P_{2}$ is the state 
$(|00\rangle +|10 \rangle +|11 \rangle )/ \sqrt{3}$.  Parameters used are listed in Table 7.}
\centerline{\footnotesize\smalllineskip
\begin{tabular}{l l l}\\
\hline
State &Desired &QNN Output\\
\hline
EPR, $\theta = 0$    &    1.0    &    1.0 \\
EPR, $\theta = \pi$   &    1.0    &    1.0  \\
Bell, $\delta = \pi$   &   1.0    &  0.99997  \\
$|00 \rangle$            &    0.0    &  $3.24 \times 10^{-5}$ \\
$|10 \rangle + 0.9 |11 \rangle$  &  0.0   &   $3.39 \times 10^{-6}$ \\
$P_{2}$                  & 0.44317   & 0.44317  \\
M                       &        0.0        & $2.59 \times 10^{-13}$ \\     
\hline
RMS   & $7.68 \times 10^{-6}$ & {} \\
\hline \hline \\
\end{tabular}}
\end{table}

The training for the partially entangled state $P$ deserves some further comment.  
We noted, above, that there is no general agreement on what the entanglement of $P$ is, though it ought to 
lie somewhere between 0 and 1 on the scale we are using.  Therefore we trained the network for various 
different target values for the entanglement of $P$, including the numbers (0.32, 0.46, 0.55) calculated 
by the three comparison methods shown in Table 5.  In Figure 1 we show the total error of the QNN for both 
the training and testing sets, as a function of the desired value.    There is a minimum at approximately 
0.44317 (though that degree of precision is probably imaginary.)  What this means is that using the value of 
0.44317 for the fourth training pair significantly increased the compatibility of the set of training and 
testing pairs, taken together; while we could train the state $P$ to any value we desire, only if we set 
that value to 0.44317 can we get at the same time the values we wish for the other training and testing pairs.  
We can say that in some sense the value of about 0.44 is the ``natural'' value for the entanglement of this state, 
at least as computed by this net; that is, using this value as the target value for $P$ leads to the greatest 
possible self-consistency for the method.  Interestingly 4/9= 0.44444… is the value for the 
entanglement of state $P$ as calculated by the formula 
$tr (\rho \tilde{\rho})$, where $\tilde{\rho}= (\sigma_{y} \otimes \sigma_{y}) \rho^{*}(\sigma_{y} \otimes \sigma_{y})$, 
which for pure states is a monotonically increasing function (like the concurrence) and thus could be used as a possible 
measure of entanglement.  This measure, however, fails for mixed states: in particular, it gives an entanglement of 1 for 
the mixed state $M$ (Table 5).  (Coincidentally the number 0.44229 for the entanglement of formation also has some special 
significance \cite{munro} for partially mixed states: this is the point at which the Werner states\cite{werner} begin to 
violate the Bell inequality.)

\begin{figure} [htbp]
%\vspace*{13pt}
\centerline{\psfig{file=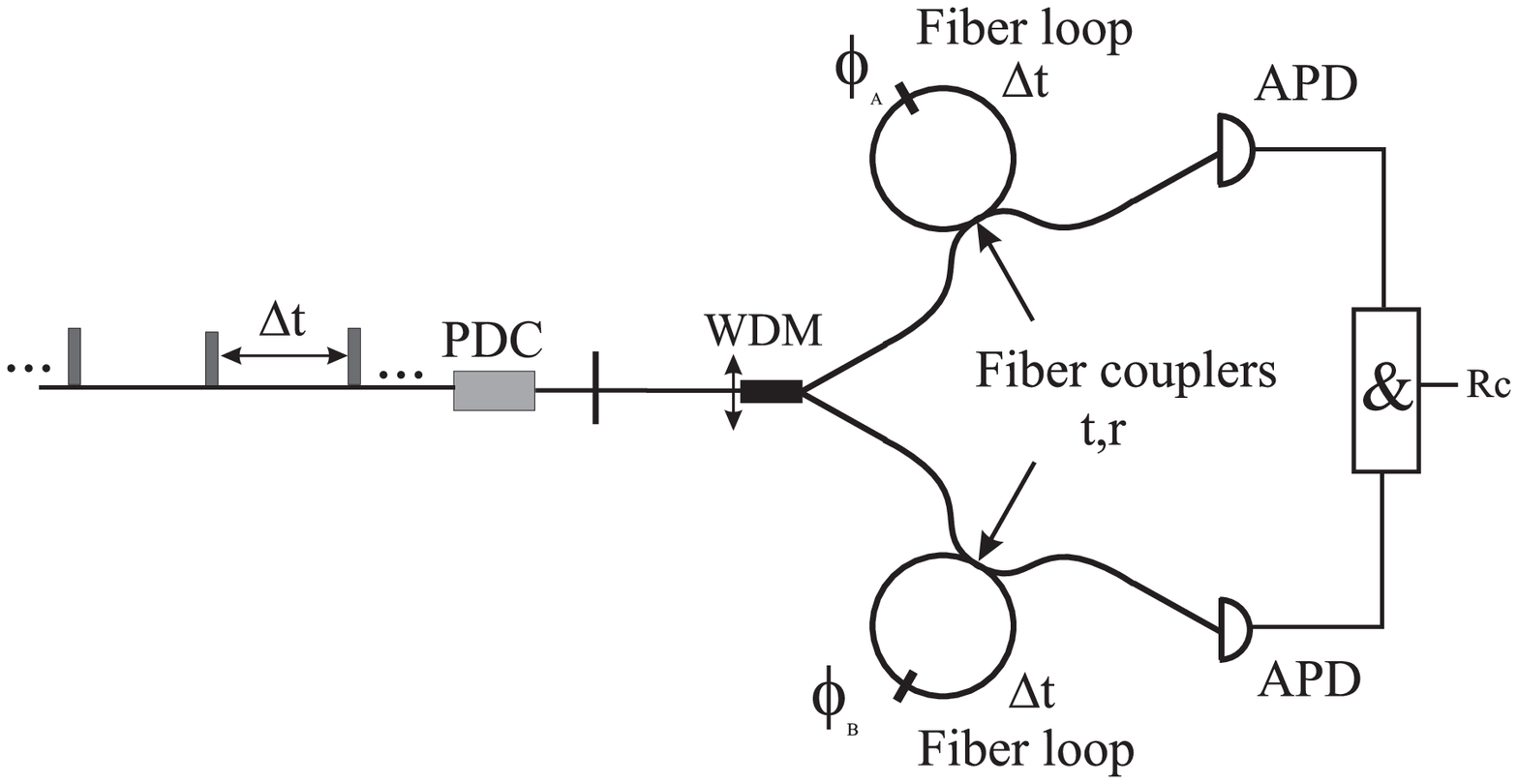, width=8.2cm, angle=90}} %100 percent
\vspace*{3pt}
\fcaption{\label{target}  
Total error, including training and testing, for different target values for the partially entangled state $P$.  
In each run, represented by a data point on the graph, all four of the training set pairs were trained, and only 
the desired value for the state $P$ was changed; all were tested on the same testing set as shown in Table 8.  
The minimum error found, $6.27 \times 10^{-6}$, occurs at a target value of 0.44317. }
\end{figure}

  It may also be of interest to note that almost all the contribution to the error comes from the partially entangled 
states in the training and testing sets. So, for example, the total error (from training and testing sets) for the 
target value of 0.32 for $P$ is about 0.2; most of this comes from the testing of state $P_{2}$, for which the net 
calculates 0.41942 instead of 0.32.  That is, except at the target value of about 0.44, the net does not recognize 
$P_{2}$ as being ``the same'' as $P$.  Results for the mixed states are almost exactly the same irrespective of the 
target value for $P$: if they were plotted in Figures 3 and 4, they would lie almost on top of the QNN values shown.  
The results for completely mixed states are always calculated to be zero ($10^{-6}$ or smaller.)
  
  Once the net is trained it is then a simple matter to calculate the output value (square of the correlation function) 
which would be measured at the given final time for any state at all, pure or mixed.  In addition to the testing set, 
we set up a grid to check that the calculated entanglement of all pure product states is zero, that is, all states of 
the form 
$(\alpha |1 \rangle + \beta |0 \rangle )_{A} (\gamma |1 \rangle + \delta |0 \rangle )_{B}/ \sqrt{\alpha \gamma + \alpha \delta + \beta \gamma + \beta \delta}$.  
The RMS error for the set of 10,000 was $1.1 \times 10^{-7}$.  Similarly we tested a grid for mixed states; the 
RMS error for the set of 10,000 was $2.6 \times 10^{-7}$.  Thus for both pure and mixed separable states the QNN gives 
reasonably good results.
  
We compare the QNN method with both a (widely-accepted) measure (Bennett and Wootters's entanglement of formation) 
and a witness (Toth and Guhne's local entanglement witness \cite{toth}, $W_{GHZn}$, which for the two-qubit system 
considered here is equal to $I- \sigma_{xA}\sigma_{xB} - \sigma_{zA} \sigma_{zB}$.)   This is because, while we 
claim to have designed only a witness, not a measure, our method turns out to be somewhat better than expected: 
to some extent, it gives information, as well, on the amount of entanglement present.    When we compare the QNN 
to the entanglement of formation, we are looking for numerical agreement ­ and get it to some extent; when we 
compare both Toth and Guhne's $W$ and the QNN, as witnesses, to the entanglement of formation, we look, as is 
appropriate, only for agreement as to whether or not entanglement is present.  Note that this last measures 
proximity to the Bell triplet state and that a negative number indicates entanglement.  
  
Figure 2 shows the QNN entanglement for the pure state 
$P_{3}(\gamma) = \frac{|00 \rangle + |11 \rangle + \gamma |01 \rangle}{\sqrt{2+|\gamma|^{2}}}$, as a function of 
$\gamma$. Values for the entanglement of formation are also shown for comparison; the results are very similar 
and constitute good agreement. The $W$ witness correctly indicates the presence of entanglement until the 
contamination gets too large: for $\gamma = 1$ $W$ is zero, which indicates no entanglement and is incorrect.  
Thus here the QNN does better as a witness than $W$.  The disagreement is probably because the $W$ witness is 
only good for states close to the Bell  triplet state $| \Phi^{+} \rangle$; for $\gamma = 1$ we are probably 
sufficiently far from the Bell triplet state that the witness no longer applies well.   
  
Figure 3 shows the calculated entanglement for the Bell triplet Werner\cite{werner} mixed states as a function 
of fidelity, again, with Bennett and Wootters's entanglement of formation and Toth and Guhne's witness $W$ for 
comparison. Again, agreement of the QNN with the former is quite good though not exact.  For fidelity between 
approximately 0.28 and 0.5 our method gives a small but nonzero entanglement, which is incorrect, since it has 
been shown\cite{bennett2} that for  $ 0.25 < F < 0.5 $ the Werner state can be written as a mixture of product 
states.  Here $W$ is a better witness than the QNN.  In Figure 4 we show the QNN entanglement for the states  
$ M'(\gamma) = \frac{\gamma|11 \rangle \langle 11| + |\Phi^{+} \rangle \langle \Phi^{+} |}{\gamma + 1}$, and 
compare those results to those for the entanglement of formation.  Again the agreement is good though not exact.  
The $W$ witness here fails for $\gamma  > 0.5$.
  
Using a larger training set -­one that includes mixed states -­does reduce the error shown in Figure 3, though 
with this model (as Figure 1 demonstrates) we have already minimized the error shown in Figure 2 by optimizing 
the target value for $P$. Changing the target value for $P$ does not alter the results for Figure 3 significantly.  
We thought it more interesting to show results for what seems to be the minimum training set for an (approximate) 
entanglement witness.
  
\begin{figure} [htbp]
%\vspace*{13pt}
\centerline{\psfig{file=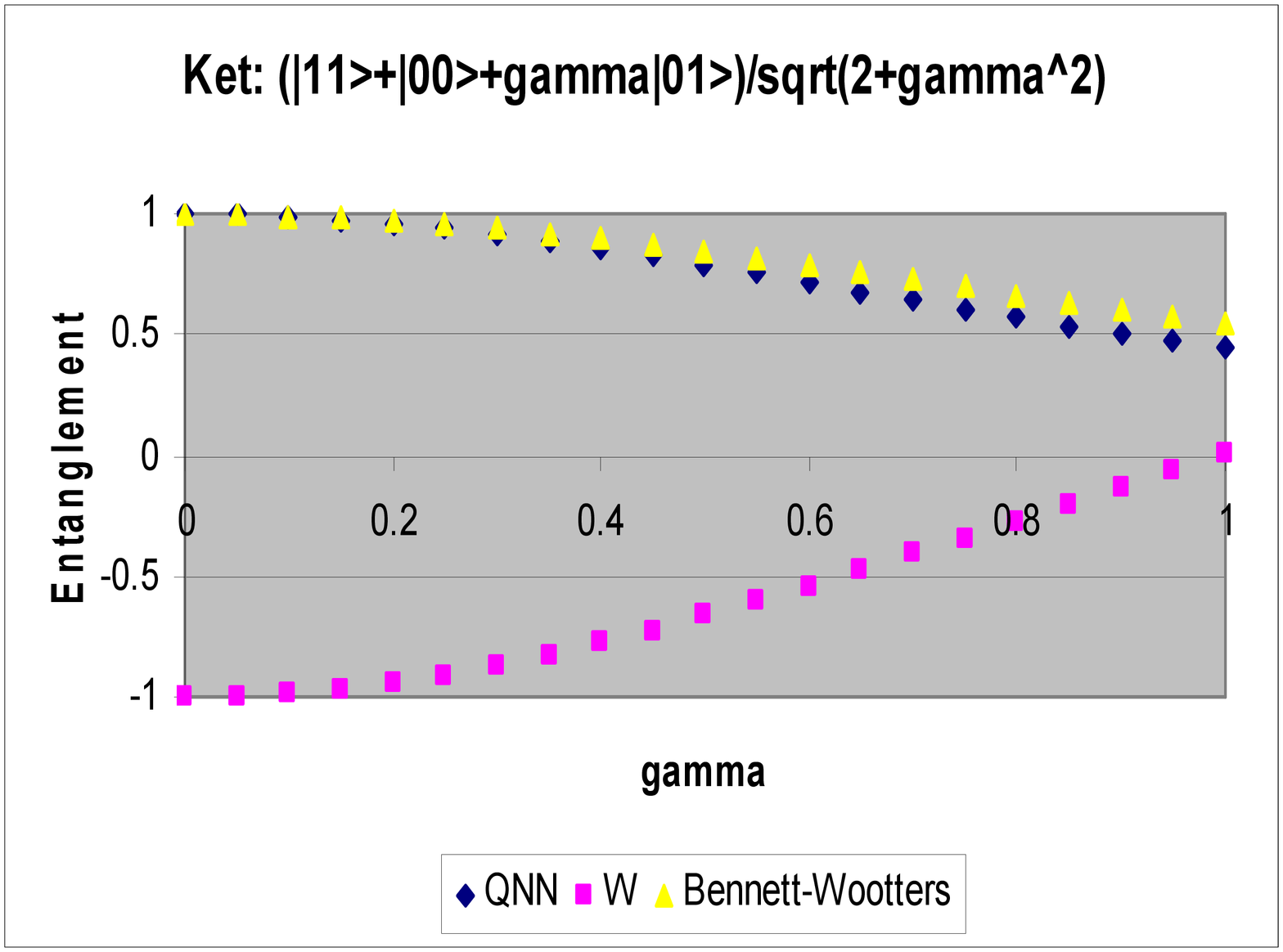, width=6.2cm, angle=270}} %100 percent
\vspace*{3pt}
\fcaption{\label{P3} Entanglement of the pure state 
$P_{3}(\gamma) = \frac{|00 \rangle + |11 \rangle + \gamma |01 \rangle}{\sqrt{2+|\gamma|^{2}}}$ , 
as a function of $\gamma$, as calculated by the QNN, using the trained parameters listed in Table 7, 
by Bennett's entanglement of formation \cite{bennett2, wootters}; and by Toth and Guhne's local 
entanglement witness $W$ \cite{toth}. Note that $W<0$ indicates entanglement.}
\end{figure}

\begin{figure} [htbp]
%\vspace*{13pt}
\centerline{\psfig{file=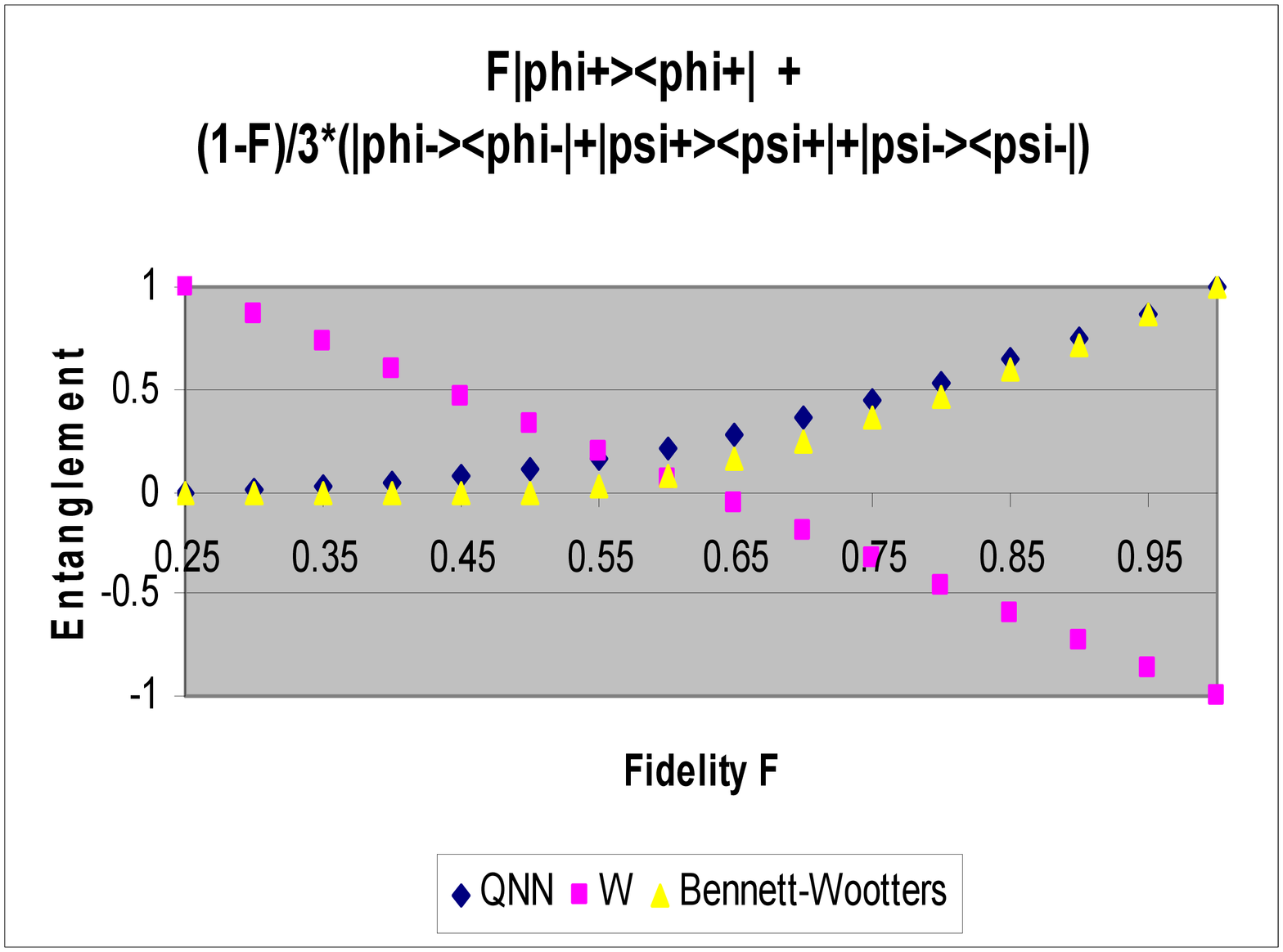, width=6.2cm, angle=270}} %100 percent
\vspace*{3pt}
\fcaption{\label{werner}
Entanglement for the Bell triplet Werner mixed states 
$F|\Phi^{+}\rangle \langle \Phi^{+}|+(1-F)/3)(|\Psi^{+}\rangle \langle \Psi^{+}| 
+|\Psi^{-} \rangle \langle \Psi^{-} |+ |\Phi^{-} \rangle \langle\Psi^{-}|)$ 
as a function of fidelity F, where 
$ |\Psi^{\pm}\rangle = (|\uparrow \downarrow \rangle \pm | \downarrow \uparrow \rangle)/ \sqrt{2}$, 
and  $|\Phi^{\pm}\rangle = (|\uparrow \uparrow \rangle \pm |\downarrow \downarrow \rangle)/\sqrt{2}$, 
as calculated by Bennett/Wootters \cite{bennett2,wootters}, by QNN, and by Toth/Guhne's $W$ witness\cite{toth}. 
Again the QNN parameters were as listed in Table 7.  The Werner states are $x=(4F-1)/3$ parts pure triplet 
(fully entangled), and $(1-x)$ parts identity operator (completely mixed)\cite{bennett2}. Note that $W<0$ 
indicates entanglement.}
\end{figure}

\begin{figure} [htbp]
%\vspace*{13pt}
\centerline{\psfig{file=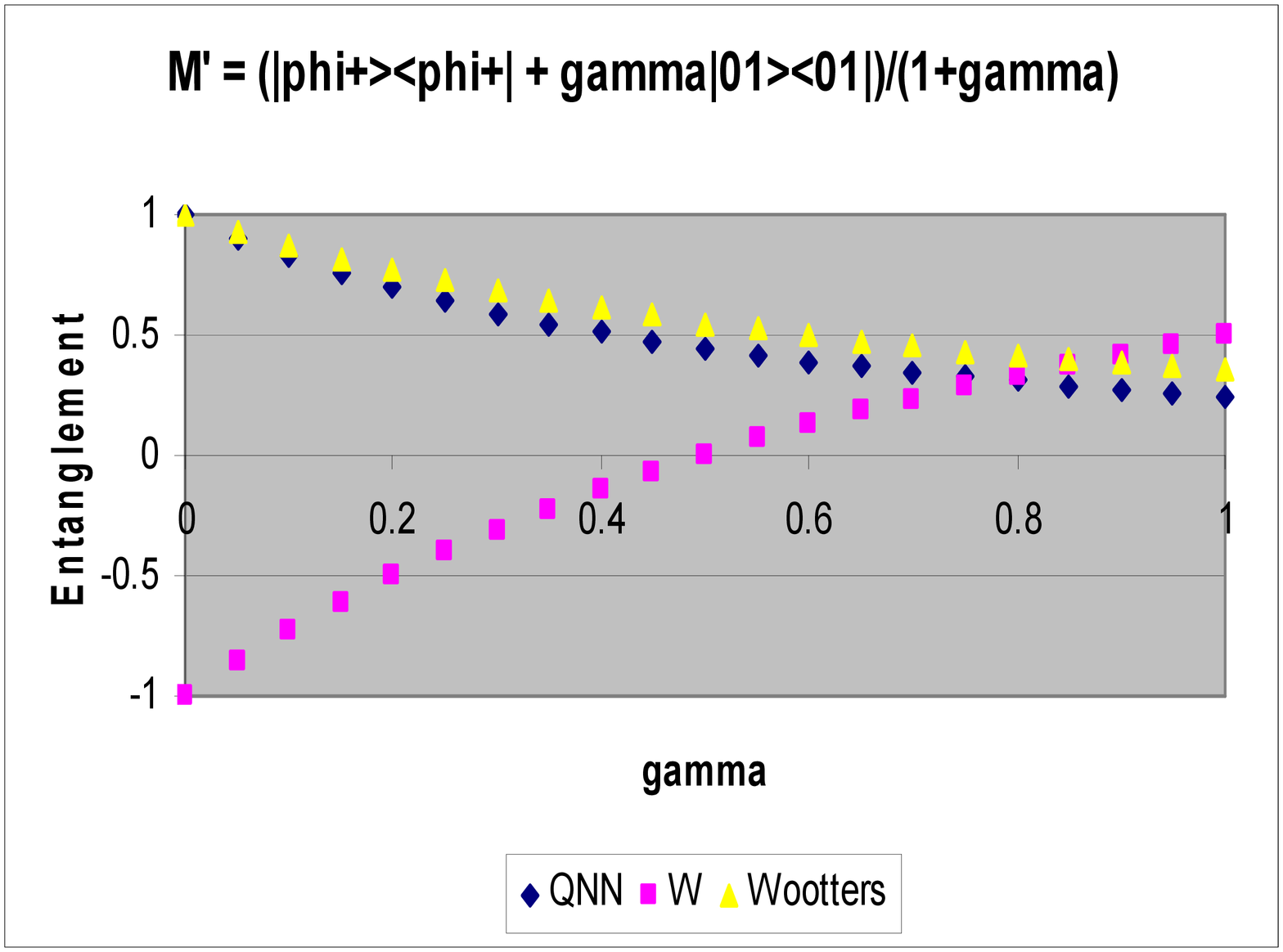, width=6.2cm, angle=270}} %100 percent
\vspace*{3pt}
\fcaption{\label{M'} Entanglement for the states 
$M'(\gamma) = (\gamma|01\rangle \langle 01|+|\Phi^{+}\rangle \langle \Phi^{+}|)/(1+\gamma)$, 
calculated by entanglement of formation,  by QNN, and by Toth/Guhne's $W$. For 
$\gamma  = 1$ this is Bennett's M state; the QNN calculates its entanglement as 0.24716, 
while Wootters's method\cite{wootters} gives 0.35458.  QNN parameters were as listed in Table 7. 
Note that $W<0$ indicates entanglement.}
\end{figure}

\vspace*{1pt}\textlineskip      %) USE THIS MEASUREMENT WHEN THERE IS
\section{ Discussion} %) A SECTION HEADING
\vspace*{-0.5pt}
\noindent

  It will have been noticed that all the coefficients on states so far trained or tested were real. 
It is a natural question at this point to ask about a phase difference, {\it e.g.}, between the two parts of a 
Bell state.  In Figure 5 we show the calculated correlation function for the Bell state 
$|00 \rangle + e^{i \theta}|11 \rangle$ as a function of initial phase difference $\theta$.  (The EPR state 
shows exactly the same dependence.)  Of course the actual entanglement is not a function of the phase difference, 
so here the QNN measure is wrong; or rather, this is the major reason our method produces only a ``witness'' not 
a ``measure.''  Interestingly Toth-Guhne's $W$ witness shows a similar oscillation, though at half the frequency. 
  
  Product states show a similar if smaller amplitude oscillation. Figure 6 shows that the oscillation in the 
product state $C$ is so small as to be negligible; $W$ correctly predicts no entanglement though we are surely 
too far from the Bell state for the method to be applicable.  Figure 7 shows the oscillation for $P_{2}$ when 
the phase difference is between the entangled pair and $|01\rangle$.  The ``correct'' target value for the $P$ 
states depends on the phase difference, and is equal to 0.44317 only for $\theta=0$; the mean value is very 
close to 4/9 (0.44444 to five significant figures, for twenty evaluated points between 0 and $2\pi$.)  Figure 8 
shows the oscillation for $P_{2}$ when the phase difference is within the entangled pair.  The oscillation from 
Figure 5 is reproduced, as expected, but on a smaller amplitude scale ( 0 to 0.44 rather than 0 to 1) because 
of the presence of the amplitude in the $|01\rangle$ state. In this case, as in Figure 5, we can see the 
oscillation in the $W$ witness as well.  Note that for this state, for any value of $\theta$, the $W$ witness 
predicts no entanglement; this is doubtless due to its being insufficiently ``close''to the Bell state.  The best 
point is at $\theta=0$ (or $2\pi$); this is the same as the point at $\gamma=1$ in Figure 2.
  
\begin{figure} [htbp]
%\vspace*{13pt}
\centerline{\psfig{file=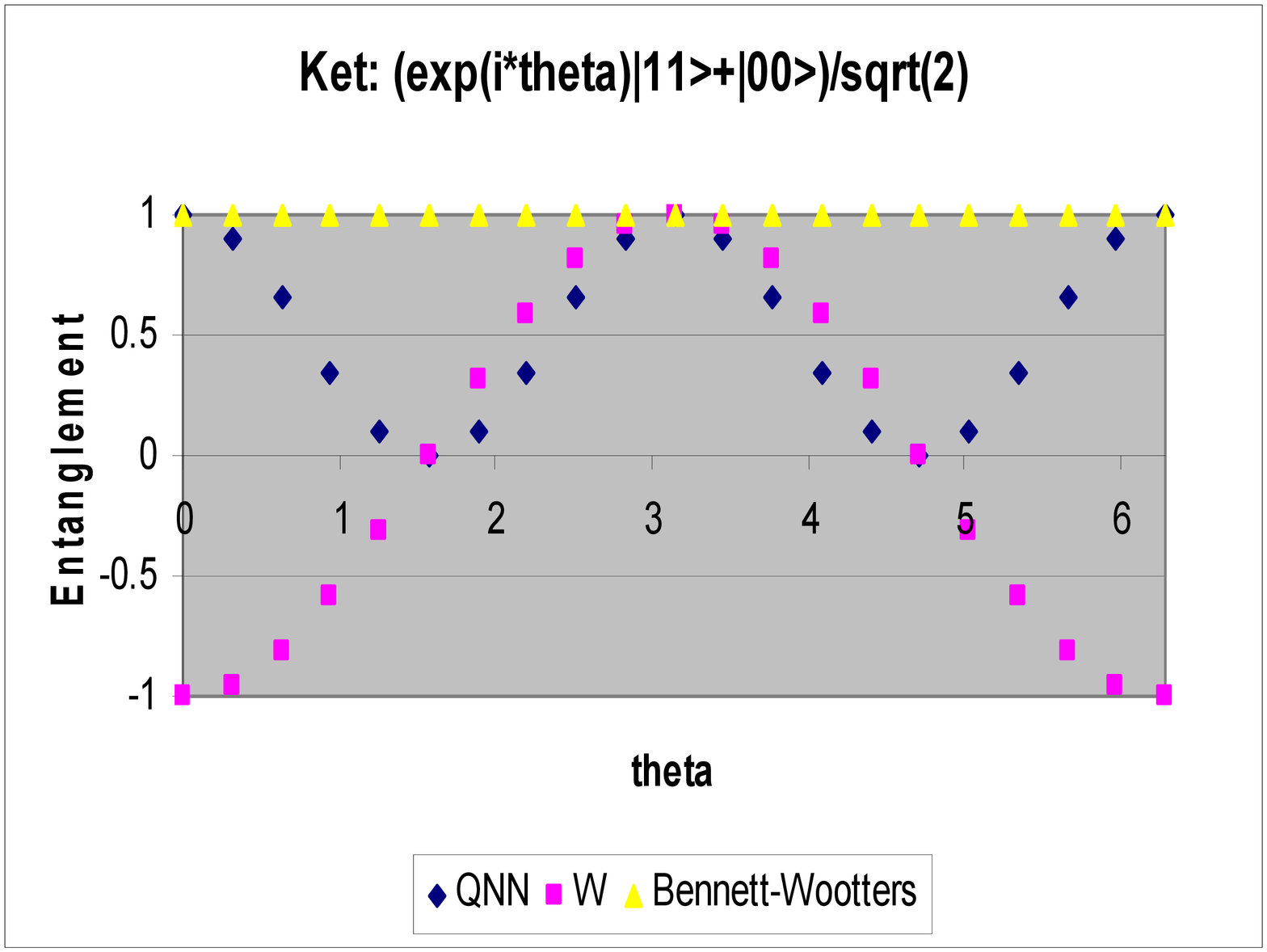, width=6.2cm, angle=270}} %100 percent
\vspace*{3pt}
\fcaption{\label{bell theta} Calculated entanglement for the state $|00 \rangle + e^{i \theta}|11 \rangle$, as a 
function of $\theta$, by Bennett-Wootters, QNN, and  Toth-Guhne's $W$.  QNN parameters were as listed in Table 7.  
Note that $W<0$ indicates entanglement.}
\end{figure}

\begin{figure} [htbp]
%\vspace*{13pt}
\centerline{\psfig{file=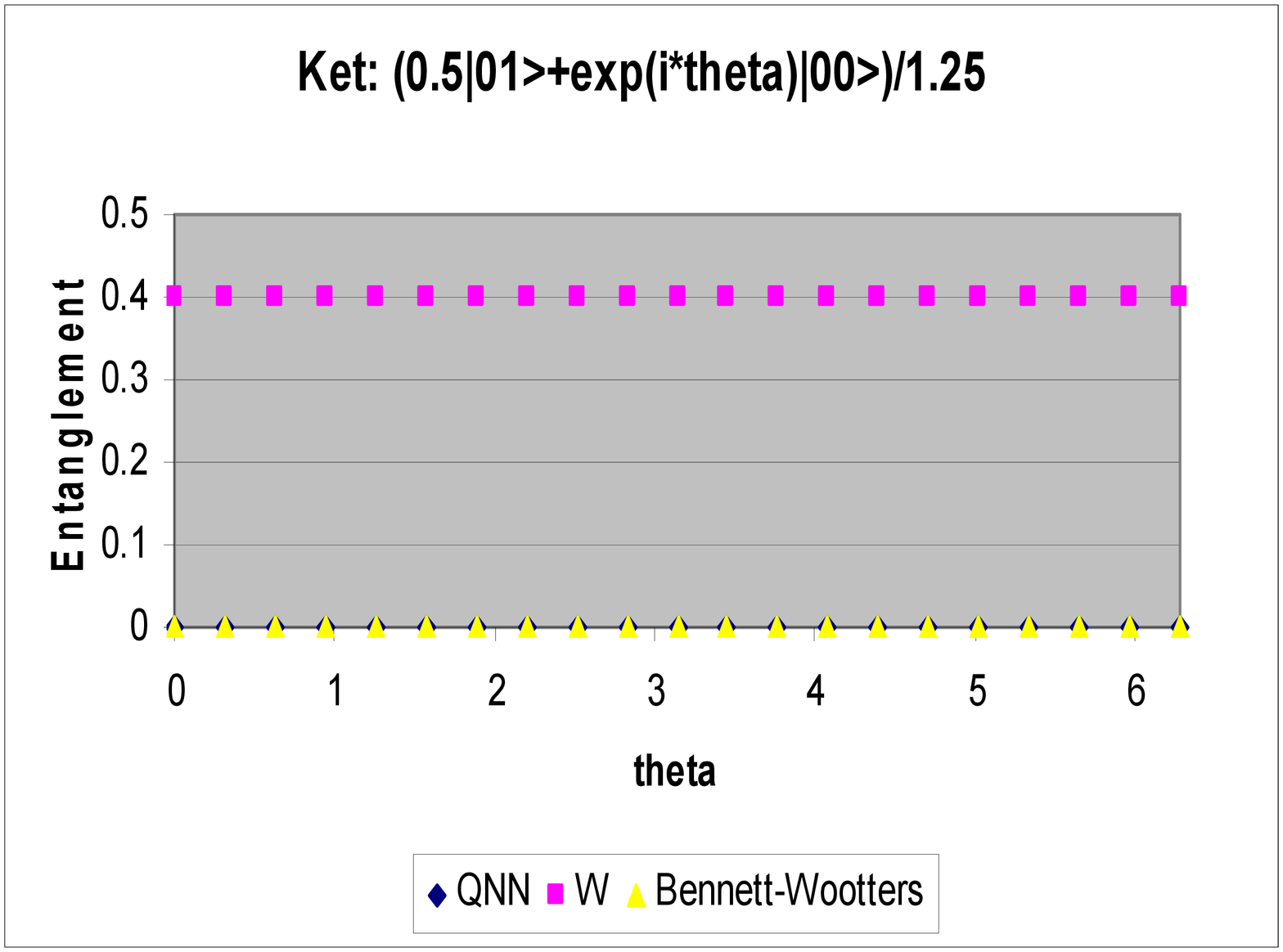, width=6.2cm, angle=270}} %100 percent
\vspace*{3pt}
\fcaption{\label{C}
Calculated entanglement for the $C$ state, by Bennett-Wootters, QNN, and Toth-Guhne's $W$.   QNN parameters were as 
listed in Table 7.  Note that $W<0$ indicates entanglement.}
\end{figure}

We can see why both witnesses behave this way by considering that any single measurement must be of the 
form of the trace of the initial density matrix times some Hermitian operator. Let that matrix be given by 
  $ \left(  \begin{array}{c c c c}
  e_{1}             & a_{1}+ib_{1} & a_{2} + ib_{2} & a_{3} + ib_{3} \\
  a_{1}-ib_{1} &        e_{2}      &  c_{1}+id_{1}   &  c_{2}+id_{2}   \\
  a_{2} - ib_{2}& c_{1}-id_{1}   &    e_{3}            & c_{3}+id_{3}   \\
  a_{3} - ib_{3}& c_{2}-id_{2}   & c_{3}-id_{3}    &   e_{4}
  \end{array} \right) $
Now, for the case of the initial state's being $|00 \rangle + e^{i \theta}|11 \rangle$, 
this means that the expectation value is given by 
$e_{1} + e_{4} + 2{\rm Re}[e^{i \theta}(a_{3} -­ib_{3})]$.  
Unless both $a_{3}$ and $b_{3}$ are zero this necessarily depends on $\theta$; even if we set them both to 
zero we still have to deal with, {\it e.g.}, the case of the state $|01 \rangle + e^{i \theta}|10 \rangle$ .  
Thus it is not possible to design a single measurement as a completely general entanglement measure.  
The oscillation that is seen in both our witness and that of Toth and Guhne is inescapable. Thus in order to 
use our method to measure independently the entanglement of an unknown state, 
it is necessary to do at least one other measurement and perhaps two.  Recently Yang and Han\cite{han2}
have devised a means of extracting an arbitrary relative phase from a multiqubit entangled state by local 
Hadamard transformations and measurements along a single basis; this method together with our own, then, 
can be used as an unambiguous entanglement witness.

\begin{figure} [htbp]
%\vspace*{13pt}
\centerline{\psfig{file=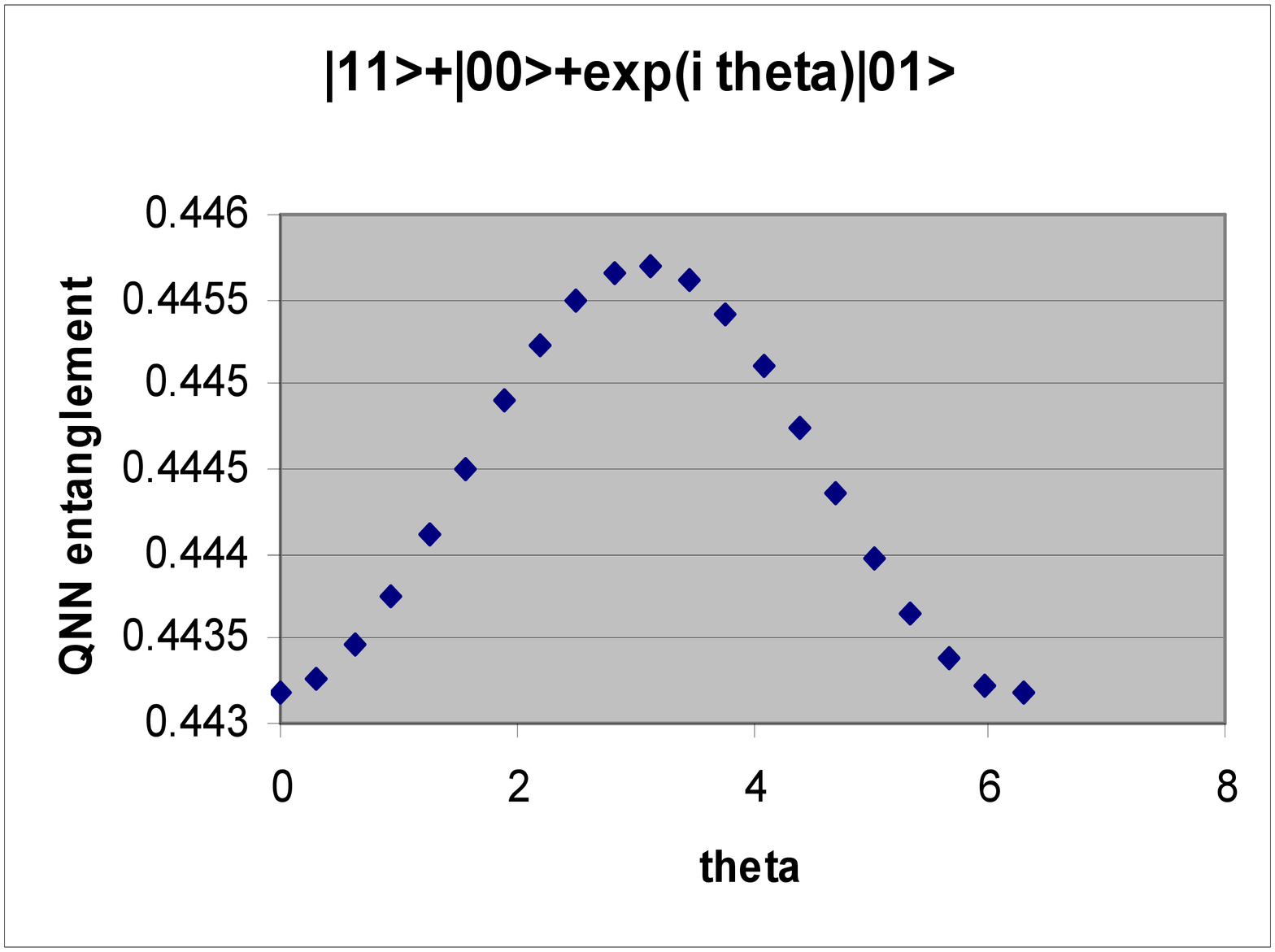, width=6.2cm, angle=270}} %100 percent
\vspace*{3pt}
\fcaption{\label{P osc}
Entanglement of the state $|00 \rangle + |11 \rangle + e^{i \theta}|01 \rangle$,  , as a function of $\theta$, as 
calculated by QNN.  QNN parameters were as listed in Table 7. Toth-Guhne's $W$ is zero for all values of $\theta$ 
(too far from the Bell state; see Figure 2.) }
\end{figure}

\begin{figure} [htbp]
%\vspace*{13pt}
\centerline{\psfig{file=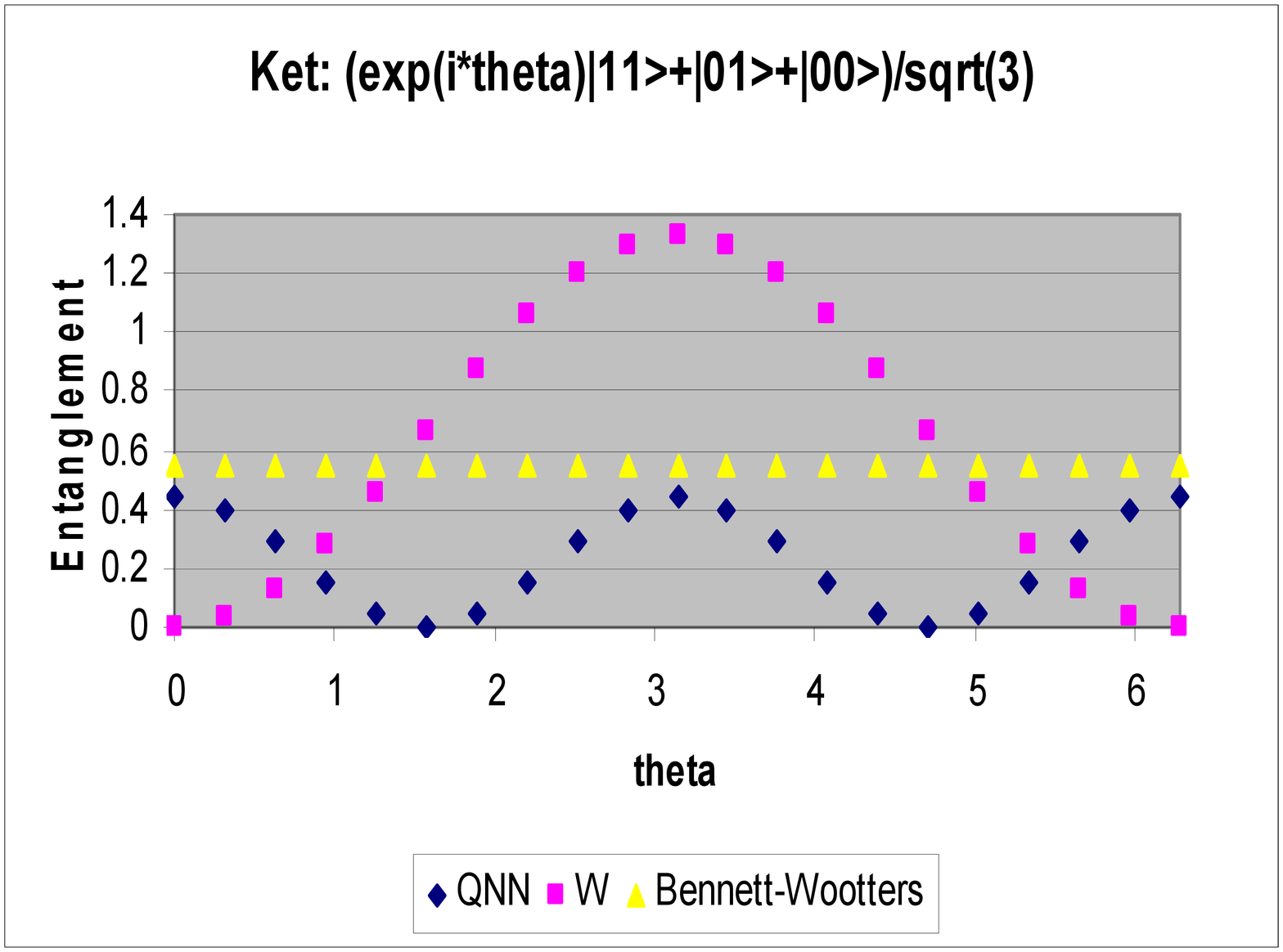, width=6.2cm, angle=270}} %100 percent
\vspace*{3pt}
\fcaption{\label{P osc2}
Entanglement of the state $|00 \rangle + e^{i \theta}|11 \rangle + |01 \rangle$, as a function of $\theta$, as 
calculated by Bennett-Wootters, QNN, and Toth-Guhne's $W$.   QNN parameters were as listed in Table 7.  Note that 
$W<0$ indicates entanglement.}
\end{figure}

One possible problem with the proposed method is that the output function chosen is always greater than or 
equal to zero; thus, in actual experimental measurements, it will systematically overestimate the entanglement 
for states close to separability.  It should also be noted that the correlation function cannot be determined 
in a single measurement, since it is an average quantity.  To measure the correlation function experimentally, 
even if the average should be very close to zero or to one, it would be necessary to produce the desired state 
many times; since in a given situation it may not be possible to do so, or without varying the phase difference, 
this may defeat the purpose.  Nevertheless we believe this approach may be a fruitful one: it is certainly easier 
to measure the correlation function than it is to determine the density matrix in full, as standard calculational 
approaches require, or even the four parameters necessary to find the concurrence\cite{horodecki}.  It is possible 
that a different, more clever choice of measurement operator(s) could reduce the number of measurements necessary 
still further.  In addition, generalization to multiple qubits or to quNits is straightforward: as long as a 
training set of sufficient size is used there is no reason to think that the net would be unable to learn the 
generalized measure.  Our experience here seems to show that the necessary size is not large.  We are currently 
pursuing these lines of research.

\vspace*{1pt}\textlineskip      %) USE THIS MEASUREMENT WHEN THERE IS
\section{ Conclusions} %) A SECTION HEADING
\vspace*{-0.5pt}
\noindent

In this paper we have developed a general dynamic learning algorithm for training a quantum computer, for 
either pure or mixed states.  We have demonstrated successful learning of some simple benchmark applications.  
We have also shown that this method can be used for the learning of an entanglement witness for an input state.  
We have shown that our witness approximately reproduces the entanglement of formation for large classes of states, 
and, while it suffers from a systematic oscillation problem, so must every single-measurement entanglement witness, 
and a method like Yang and Han's can be used in conjunction to take care of the problem.  It is superior to other 
witnesses in that, first, the state need not be ``close'' to a particular kind of entangled state ­or, indeed, to 
any particular state; and, second, that the state may be completely unknown.  Generalization to systems of more than 
two qubits and to multiple level systems is in progress.

\nonumsection{Acknowledgements}
\noindent
  This work was supported by the National Science Foundation, Grants ECS-9820606 and 0201995.  We thank W.K. 
Wootters, W.J. Axmann, and A.J. Schauf for helpful discussions.

\nonumsection{References}

\end{document}